\documentstyle[11pt,aaspp4, flushrt]{article}  

\begin{document}
 
\title{Spectral Evolution of Circinus X--1 Along its Orbit}

\author{R. Iaria\altaffilmark{1}, T. Di Salvo\altaffilmark{2},
L. Burderi\altaffilmark{3}, N. R. Robba\altaffilmark{1}}
\altaffiltext{1}{Dipartimento di Scienze Fisiche ed Astronomiche, 
Universit\`a di Palermo, via Archirafi n.36, 90123 Palermo, Italy}
\authoremail{iaria@gifco.fisica.unipa.it} 
\altaffiltext{2}{Astronomical Institute "Anton Pannekoek," University of 
Amsterdam and Center for High-Energy Astrophysics,
Kruislaan 403, NL 1098 SJ Amsterdam, the Netherlands.}
\altaffiltext{3}{Osservatorio Astronomico di Roma, Via Frascati 33, 
00040 Monteporzio Catone (Roma), Italy}

\begin{abstract}

We report on the spectral analysis of Circinus X--1 
observed by the ASCA satellite in March 1998 along one orbital period.
The luminosity of the source (in the 0.1--100 keV band) ranges from
$2.5 \times 10^{38}$ erg s$^{-1}$ at the periastron (orbital phase $0.01$)
to $1.5 \times 10^{38}$ erg s$^{-1}$ at orbital phase $0.3$.
From the spectral analysis and  the lightcurve we argue
 that  Cir X--1 shows three states along the orbital evolution.   
The first state is at the orbital phase interval 0.97--0.3: 
the luminosity becames
super--Eddington and a strong flaring activity is present.  In this
state a shock could form in the inner region of the system due to
the super--Eddington accretion rate, producing an outflow of ionized matter
whose observational signature could be the prominent absorption edge at 
$\sim 8.7$ keV observed in the energy spectrum at these phases. 
In the second state, corresponding to the orbital phase interval between
0.3 and 0.7, the accretion rate is sub--Eddington and we observe a weaker
outflow, with smaller hydrogen column: the absorption edge is now at 
$\sim 8.3$ keV with an optical depth a factor of 2.5 to 6 smaller.
The third state corresponds to the orbital phase interval 0.78--0.97.
In this state the best fit model to the spectrum requires the presence
of a partial covering component, indicating that 
the emission from the compact object is partially absorbed by neutral 
matter, probably the atmosphere of the companion star and/or the accreting 
matter from the companion. 
\end{abstract}

\keywords{stars: individual: Cir X--1 --- stars: neutron stars ---
X-ray: stars --- X-ray: spectrum --- X-ray: general}

\section{Introduction}
Circinus X--1 (Cir X--1) is a puzzling X--ray binary.  
Because of its rapid
variability a black hole was supposed to be the compact object in this
system (Toor, 1977), until type--I X--ray bursts were detected (Tennant
et {\it al.}, 1986 a,b), suggesting that the compact object is a
neutron star. A modulation with a period of 16.6 days was observed at 
several wavelengths in the emission of the source. The high stability of 
this period was a strong indication
that it is the orbital period of Cir X--1.  The orbit is highly
eccentric ($e \sim 0.9 $) (Murdin et al. 1980; Tauris et al. 1999).
Periodic radio flares near the zero phase (i.e.\ at the periastron, for
recent ephemeris see Stewart et {\it al.}, 1991) are also accompanied
by drastic changes in the X--ray lightcurve (X--ray flares and dips).
The position of the probable radio counterpart is close to the
supernova remnant G321.9--0.3 (Clark, Parkinson \& Caswell, 1975),
suggesting that Cir X--1 could be a runaway binary. Case \&
Bhattacharya (1998) have revised the distance to the supernova remnant
G321.9-0.3, and hence to the associated Cir X--1, to 5.5 kpc. Therefore
in the following we adopt 5.5 kpc as the distance to the source.  

Quasi periodic oscillations (QPOs) at 100--200 Hz were observed in this
source (Tennant, 1987; Shirey et al., 1996), but no kHz QPOs were
detected.  Recently Iaria et al. (2001) detected the presence of highly 
ionized matter around Cir X--1, with hydrogen column density of $\sim 10^{24}$ 
cm$^{-2}$, at the orbital phase 0.11--0.16. The continuum emission
was fitted using a Comptonization model plus a power--law tail at
energies higher than 10 keV. The luminosity of the power law is $4 \%$
of the total.  Brandt \& Schulz (2000), analysing Chandra observations taken 
at the periastron, detected, for the first time in the X-ray band, the
presence of P--Cygni lines in X--ray band. They explain these features
as due to the presence of high velocity outflows from this system 
(probably a moderate temperature, $\sim 0.5$~keV, wind from the X-ray 
heated accretion disk).
Johnston et al. (1999), using several infrared (IR) and
optical data from 1976 to 1997, observed along the whole orbit an
asymmetric H${\alpha}$ emission line.  They argued that the broad
component of the emission line arises in a high--velocity ($\sim 2000$
km s$^{-1}$), optically thick, flow near the neutron star.  Brandt et
al. (1996), using ASCA data taken on 1994 August 4-5, 
found a sudden variation of the flux at
the periastron phase, the count rate increasing from $\sim 30$ counts
s$^{-1}$ to $\sim 300$ counts s$^{-1}$. In the low-count rate state 
they obtained a good fit using a partial covering component, in addition 
to a two--blackbody model, with a corresponding hydrogen column of $\sim
10^{24}$ cm$^{-2}$; in the high-count rate state the partial covering 
of the X-ray spectrum was not needed. Because of the similarity with 
the spectra of Seyfert 2 galaxies with Compton-thin tori, Brandt et al. 
(1996) suggested that matter at the outer edge of the accretion 
disk, together with an edge-on disc orientation, could explain the partial 
covering of the spectrum in Cir X--1.

Here we report the results of the spectral analysis in the energy
range 0.6--10 keV performed on Cir X--1 data from the ASCA satellite
taken in 1998. These data cover a whole orbital period allowing to study the
corresponding changes in the properties of the matter around the system.

\section{Observations and Lightcurve}

We performed a systematic spectral study of Cir X--1 along its orbital
period using ASCA archival data taken in 1998 March between 3 and 18. 
The instruments on board ASCA
(Tanaka, Inoue, \& Holt, 1994) consist of two solid--state imaging
spectrometers (SIS) and two gas imaging spectrometers (GIS, Ohashi et
al., 1996; Makishima et al., 1996).  In our analysis we used only data
from GIS instruments. GIS have an energy resolution of $\sim 8\%$
(FWHM) at 6 keV and their detection efficiency above $\sim 3$ keV is
higher than with SIS (it is twice as large as the SIS detection efficiency 
at energies of 6.5--7 keV).  
We used data at high telemetry from the ASCA archive,
screened using current standard filter criteria. The GIS photon extraction 
region was limited to the standard circular region of $6'$ radius
centered on the source. The background was taken from a $6'$ radius
circle free of emission from Cir X--1 in the external region of the
GIS field of view and after an accurate investigation of the uniform
background distribution. Because of the high luminosity of the source
it was necessary a deadtime correction after the extraction of the
lightcurves and the spectra. To correct for this effect we used the ftools 
{\it ldeadtime} and {\it deadtime} for lightcurves and spectra, respectively,
using the corresponding mkf files (which contain the time-histories of 
various parameters from which good data can be identified and screened) 
for each observation.  A systematic error of $2\%$ was added to the spectra.  
The observations used for the analysis cover an entire orbital period  
from 1998 March 3 to March 19. In Table 1 we report for each of the seven 
analysed observations (labelled A, B,..., G) the start and stop times, the
exposure time and the relative orbital phase assuming the orbital
ephemeris as reported by Stewart et al. (1991).

In Figure 1 we plotted the Cir X--1 lightcurve in the energy band
0.6--10 keV (upper panel), and the corresponding hardness ratio, i.e.\
the ratio of the count rate in the energy band 3--7 keV to that in the 
band 1--3 keV (lower panel).  In the lightcurve large variabilities 
are observed, especially at orbital phase 0--0.2, where sudden 
flux variations are visible, with the count rate ranging
from $\sim 400$ counts s$^{-1}$ to $\sim 600$ counts s$^{-1}$. In the
same phase interval we observe a rapid decrease of the hardness ratio
from $\sim 0.95$ to $\sim 0.4$.  In the phase interval 0.3--0.6
the lightcurve does not show any significant variability and the count
rate is $\sim 300$ counts s$^{-1}$, while the hardness ratio increases
from 0.7 to 0.9. In the phase interval 0.75--0.88 the lightcurve shows
a count rate slightly higher than 200 counts s$^{-1}$ and the hardness
ratio is constant at $\sim 1.05$.  The longest observation on 
March 3 shows a large variability in the hardness ratio. Therefore
we divided this observation into four intervals (labelled A1, A2, A3 
and A4), the
first corresponding to the orbital phase 0.95, the second to 0.98--1,
the third to 0.01--0.04 and the fourth to 0.05--0.09.  In Table 2 we
report the intervals in which the lightcurve was divided and the
corresponding orbital phases. We also show for each of the intervals 
the flux of the source extrapolated in
the 2.5--25 keV energy range, in order to compare
these values with previous observations. The flux in the band 2.5--25
keV is very similar to that measured by Shirey et al. (1999) with the
RXTE satellite.  In the same table we also report the flux of the source 
extrapolated in the 0.1--100 keV energy range (using the best
fit spectral model discussed in the next section), and the
corresponding luminosity of the source.

\section{Spectral Analysis}
  
For each of the ten intervals described above, corresponding to
different orbital phases, we extracted energy spectra, in the band
0.6--10 keV, in order to study the spectral variations along the orbit.  
The continuum emission was fitted using a Comptonization model 
({\it Comptt}, Titarchuk 1994), modified, at low energies, by 
photoelectric absorption by cold matter.  
In a previous paper (Iaria et al., 2001) the broad band (0.1--100 keV)
spectrum from the BeppoSAX satellite did not allow to distinguish
between a simple Comptonization and the so-called Eastern model,
consisting of a multicolor disk blackbody plus a blackbody component
(Mitsuda et al., 1984). However, because of the large value of the
neutron star radius obtained from the fit ($\sim 35$ km), we preferred
the Comptonization model to fit the spectrum.  Therefore we use the
same model here.  The average photoelectric absorption, 
obtained from the fit, is $\sim 1.6 \times 10^{22}$ cm$^{-2}$. 
The seed photon temperature varies from 0.51 keV
near the periastron to 0.37 keV at orbital phase 0.89.
The electron temperature varies from 1 keV (at orbital phase $\sim
0.2$) up to 1.7 keV (at orbital phase 0.89). 
The Compton optical depth varies between 16 and
20 along the orbital period.  

In spectra A1, F, and G we improved the
fit adding a partial covering to the model. The probability of chance
improvement for the addition of this component, estimated using an 
F-test, was 0.08, $2.8 \times 10^{-10}$ and $5.3 \times 10^{-5}$,
respectively. Note that this component is essentially needed to fit
an iron absorption edge. In fact we obtain equivalently good fits
if we use an absorption edge, at energies between 7.6 and 7.9 keV, 
instead of the partial covering for these intervals. In this case,
the edge optical depth ranges from 0.008 for interval A1 to 0.12 for
interval F.  However, in agreement with the results of Brandt et al. 
(1996), in the following we adopt and discuss the partial covering 
interpretation.

In some spectra (A4 to E) we detected the presence of an
absorption edge from highly ionized iron.  In the spectra where the
partial covering is present there is not evidence of a such feature.
In A2, A3 and D neither the partial covering nor the absorption edge are
present; in these cases we could only find an upper limit on the 
optical depth of the absorption edge.
An emission line from ionized iron is also needed to fit some of the
spectra (A2 to A4 and C to E).  In Table 2
we show the spectral features that were included in the best 
fit model for each of the ten spectra, and in Tables 3 and 4 we 
report the parameters of the best fit model, together with the 
probability of chance improvement of the fit when the indicated
component is included in the spectral model.

\section{Discussion}

We analyzed data of Cir X--1 from seven ASCA observations (spanning an
entire orbital period of the X-ray source) in the energy range 0.6--10 keV.  
From these observations we extracted ten spectra corresponding to 
different orbital phases.  The lightcurve shows large variability at
the orbital phase interval 0--0.2.  At phases 0.3--0.6 the lightcurve
is constant and at phases 0.78--0.89 the count rate decreases.

The equivalent hydrogen absorption column, $N_H$, derived from the 
best fit model to these spectra is $ \sim (1.6-1.7)
\times 10^{22}$ cm$^{-2}$, in agreement with previous observations
(Brandt et {\it al.}, 1996; Iaria et al., 2001). For a distance to the
source of 5.5 kpc the visual extinction in the direction of Cir X--1
is $A_v=5.8 \pm 2.0$ mag (Hakkila et al. 1997).  Using the observed
correlation between visual extinction and absorption column (Predehl
\& Schmitt 1995) we find that the expected hydrogen column for 
Cir X--1 is $N_H=\left(1.38 \pm 0.02 \right) \times
10^{22}$ cm$^{-2}$.  It is slightly smaller than the value obtained
from the fit, probably because of the presence of obscuring matter
close to the X--ray source.  The small fluctuations of the measured
Galactic absorption along the orbital phase, which seem to be statistically
significant for the intervals A3 and A4 (see Tab.~3), can be caused by 
fluctuations in the density (and/or in the ionization level) of this 
excess of matter around the system. 

We fitted the continuum using a
Comptonization model.  The values of the seed--photon and
electron temperatures with respect to the orbital phase are plotted
in Figure 2 (upper and middle panel, respectively).  The spectra A1, F
and G require in addition a partial covering component to fit an
iron edge at 7.6--7.9 keV.  An absorption
edge from ionized iron is present in spectra A4, B, C and E.  In
spectrum D the addition of this component is not statistically 
significant and we find only an upper limit for its optical depth.  
The unabsorbed flux,
extrapolated in the band 0.1--100 keV, is reported in Figure 3. In
this figure we marked the corresponding Eddington flux for a neutron
star of $1.4 M_{\odot}$ at a distance of 5.5 kpc. The flux seems to be
super--Eddington at the orbital phases 0--0.2 and 0.95--1 (near the
periastron phase).

For each spectrum we calculated the radius, $R_W$, of the seed-photon
Wien spectrum using the parameters reported in Tables 3 and
4. Assuming a spherical geometry, this radius can be
expressed as $R_W=3 \times 10^4 D \sqrt \frac{f_{bol}}{1+y}/
\left(kT_0 \right)^2$ km (in 't Zand et {\it al.} 1999), 
where $D$ is the distance to the source in
kpc, $f_{bol}$ is the unabsorbed flux of the Comptonization spectrum in 
erg cm$^{-2}$ s$^{-1}$, $kT_0$ is the seed--photon temperature in keV, 
and $y=4kT_e\tau^2/m_ec^2$ is the relative energy gain due to the
Comptonization.  The obtained radii, which are reported in Tables 
3 and 4, are in the range between 60 and 120 km.
For the sake of clarity $R_W$ is also plotted versus the orbital phase
in Figure 2 (lower panel).  Following Iaria et al.  (2001) we
interpret this radius as the inner radius of the disk.  Tennant
(1987), using EXOSAT data, and Shirey et al. (1996), using RXTE data, 
observed the presence of a QPO at frequencies between 100 and 200 Hz
in Cir X--1.  This QPO has been tentatively identified with the lower
kHz QPO based on the observed correlations among the QPO frequencies 
in NS LMXBs and black hole binary systems (Psaltis, Belloni, \& 
van der Klis 1999), although the interpretation of this feature needs 
to be confirmed. 
However, we observe that assuming that the frequencies of this QPO
are related to the Keplerian velocity of the plasma in the
accretion disk, we deduce that the corresponding radii are between 60
and 120 km, similar to the obtained radius of the seed--photon emitting
region. Note also that the other luminous Low Mass X--ray Binaries (LMXBs) 
of the Z--class show a radius of the seed--photon emitting region much 
closer to the neutron star radius as compared to Cir X--1 (see Di Salvo 
et al., 2000; Di Salvo et al., 2001), and, accordingly, these Z sources 
present QPOs at kHz frequencies in their power spectra.  Note, however,
that interpreting the seed-photon radius as the inner radius of the disk
would imply a quite high accretion rate through the disk of $\sim 3 \times
10^{18}$ g/s, i.e.\ $\sim 3$ times the Eddington accretion rate, for
$R_W = 120$ km and $kT_0 = 0.4$ keV.

A possibility to explain the large seed--photon radius in
Cir X--1 is the presence of a relatively large magnetic field: we
obtain a strength of the magnetic field in Cir~X--1 of $\sim 10^{10}$ G,
assuming that the measured seed-photon radius is the magnetospheric
radius. This strength is one order of magnitude higher than 
the supposed magnetic field for LMXBs of the Z class.  Moreover, the
interpretation of $R_W$ as the magnetospheric radius would imply lower
values of $R_W$ at higher flux levels, that is not observed (see
Fig.~3), although the error bars are large especially at phases
between 0.3 and 0.9.

Another possibility to explain the large value of the seed--photon
radius is that the inner region of the disk is not directly visible, for the
presence of optically thick material in the central part of the
system, such as the optically thick corona that probably produces the
Comptonization spectrum, or it is absent, because, for instance, its matter
is expelled in a jet or wind.
We observe that the seed--photon radius seems to increase during the flaring
activity at the orbital phase between 0 and 0.2. We could suppose that
because of the high (probably super--Eddington) accretion rate, a shock is 
produced in the 
inner region of the system, pushing the accreting matter towards the external 
regions.  A similar scenario was proposed by Haynes et al. (1980), who 
explained  the observed radio flares at the periastron as coming from a
shock front formed by the super--Eddington accretion.
Also, as Cir~X--1 is thought to be observed at a high inclination
(Brandt et al. 1996),
it is possible that the optically thick material at the shock
intercepts and reprocesses the radiation from the inner regions, which
can therefore not be directly observed.

The Comptonized spectrum could originate in an accretion disk corona
(ADC) that could be formed by evaporation of the outer layers of the
disk illuminated by the emission of the central object (White \& Holt,
1982).  The radius of the corona can be written as $R_{cor} \simeq
\left(M_{NS}/M_{\odot} \right) T_7^{-1} R_{\odot}$ (White \& Holt,
1982), where $M_{NS}$ is the mass of the compact object, $M_{\odot}$
and $R_{\odot}$ are mass and radius of the Sun, and $T_7$ is the ADC
temperature in units of $10^7$ K. Under this hypothesis, using the
values reported in Tables 3 and 4, we find that the radius of the ADC,
$R_{cor}$, varies along the orbit in the interval $(4.8-7.8) \times 10^5$ km.
We can infer the density of the ADC using the relation $\tau =
\sigma_T N_e R_{cor}$, where $\tau$ is the optical depth obtained from
the fit, $\sigma_T$ is the Thomson cross-section, $N_e$ is the number
of particles per unit volume and $R_{cor}$ is the ADC radius
calculated above.  Note that we are considering $N_e$ constant along
the radius of the corona, that is a rough approximation. Under this
hypothesis $N_e$ varies between $(3.1-6.3) \times 10^{14}$ cm$^{-3}$.
These are typical values of density for an ADC (Vrtilek et al., 1993).  

We observe that the electron temperature increases in the orbital 
phase interval 0.2--0.9 while it decreases in 0.9--1 and 0--0.1. 
This behavior can be due to the fact that at orbital
phase 0.9 the the accretion rate increases
and consequently the inner temperature of the accretion disk increases 
(as is seen in Fig. 2, upper panel).
The injection of more soft photons in the corona enhances the Compton 
cooling and the electron temperature decreases. 
In the orbital interval 0--0.2 the radiation
pressure moves the shock to $\sim 120$ km, the seed--photon
temperature decreases, and so does the electron temperature.
After, at orbital interval 0.2--0.9, the injection
of soft photons decreases, the cooling of the corona becames less
efficient and the electron temperature increases.

An iron emission line at energies 6.6--6.8 keV is present in some
spectra (A2, A3, A4, C, D and E), corresponding to emission from 
Fe XXIII--XXV.
The equivalent width of the line is $\sim 30-70$ eV.  
We have investigated the hypothesis that this emission line
could be produced by reprocessing of X--rays, coming from the corona
or the inner part of the system, by the disk surface (e.g.\ Brandt
\& Matt, 1994).  If Cir X--1 is observed at high inclination 
(as suggested by Brandt et al. 1996) an iron emission line at
$\sim 6.7$ keV could be produced by the mechanism described above,
but the corresponding equivalent width should be $\sim 20$ eV and the
width of the line, $\sigma_{Fe}$, between 0.6--0.8 keV (see Fig. 2,
3, and 4 in Brandt \& Matt, 1994). The predicted equivalent width
seems too low to explain the observed line in Cir X--1.  Therefore,
in agreement with Chandra results (Brandt \& Schulz 2000) we will
consider in the following that the iron emission line is produced in 
a outflow of ionized matter from the accretion disk (see below).

\subsection{The Scenario in Cir X--1} 

In this section we summarize our interpretation of the spectral evolution 
of Cir X--1 along the orbit based on the results from the analysis 
described above, as well as the BeppoSAX results (Iaria et al., 2001), the
Chandra results (Brandt \& Schulz, 2000) and the detection in IR and 
optical bands of emission lines 
along the orbital period of the source (Johnston et al., 1999).

The broader component of the H${\alpha}$ emission line, observed
along the whole orbit of the source, can be interpreted as produced
in an outflow of matter with a velocity of $\sim 2000$ km s$^{-1}$
(Johnston et al., 1999).  This detection gives us the information that
an outflow of matter, from the compact system, is present in Cir
X--1. The lack of the red wing of the line is associated to an
optically thick outflow.  Chandra satellite observed, near the
periastron (phase $\sim 0.99$), the presence of discrete features
with P--Cygni profiles in the X--ray band (Brandt \& Schulz, 2000). 
The inferred velocity of the outflow is again $\sim 2000$ km s$^{-1}$, 
in agreement with the results of the IR observations. 
Moreover, the fact that P--Cygni profiles are observed for Fe XXIV 
and other highly ionized elements, suggests that the matter in the 
outflow at the periastron phase is highly ionized. 
Iaria et al. (2001) observed near the periastron (phase $\sim 0.11$) 
the presence of an absorption edge at energy $\sim 8.7$ keV  
corresponding to highly ionized iron. Brandt \& Schulz (2000)
argue that the absorption edge could be associated to the outflow,
probably consisting of a thermally driven wind formed by the 
X--rays from the inner region which heat the disk surface. 
It is not straightforward to infer the geometry (radial or vertical)
of the outflow from these data. However, we can see that the model of 
thermally driven wind proposed by Begelman et al. (1983) could explain 
the observed outflow.  In this model the outflow is near the disk surface and
mainly in the direction of the observer 
(Begelman et al. 1983; Begelman \& McKee 1983). This seems reasonable 
in the case of Cir X--1 given that Brandt et al. (1996) argued that 
Cir X--1 is almost edge--on, and the P-Cygni profiles (to date 
the only ones observed in a LMXB) indicate that we are observing both the 
approaching and the receding part of the outflow.  This geometry could 
also explain the large optical depth of the iron absorption edge 
($\tau \sim 1$; Iaria et al., 2001).  However, the model of thermally 
driven wind does not take into account the radiation pressure, and
is valid for luminosities of a few percent of the Eddington
luminosity or less (Begelman et al. 1983), while 
the radiation pressure could be significant in Cir X--1 because of the 
high luminosity of the source, especially near the periastron.

The study of Cir X--1 along the entire orbit allows to describe the
evolution of the iron absorption edge and emission line present in the
spectra with a scenario where an outflow from the system is almost
always present, but its physical conditions change as function of the
accretion rate along the orbital period.  We distinguish three states
of the source along the orbit:
\begin{enumerate}
\item A super--Eddington state at phases 0.99--1 and 0--0.20, when the 
outflow of highly ionized matter (and probably a shock too) is formed;
\item A quiet state between 0.3--0.7, when the source has a sub--Eddington 
luminosity and the outflow is weaker;
\item A partial covering state between 0.77--0.95, when the atmosphere
of the companion star and/or the accreting matter from the companion,
which probably forms a bulge in the outer disk, partially cover the 
emission from the compact object.
\end{enumerate}

\subsubsection{The super--Eddington state: presence of an outflow of
highly ionized matter}

The spectra A2, A3, A4 and B belong to the same state of the source,
corresponding to the phase interval around the periastron, 0.99--1 
and 0--0.2. In these
spectra there is not evidence of partial covering, suggesting the
absence of neutral matter around the system (contrarily to what is
observed at previous orbital phases, between 0.78--0.96, see below).  
The flux is super--Eddington for each spectrum, for the assumed
distance and in the hypothesis that the emission from the source
is mainly isotropic .  In this state probably a
shock forms in the inner region of the system, and radio and X--ray
flares might be associated to the shock front (Haynes et al.  1980).
The P--Cygni profiles observed by Brandt \& Schulz (2000) could be
produced by the high luminosity (and then the high radiation pressure)
that sweeps away and ionizes the accreting matter around the system.
In spectra A2 and A3 we do not detect the absorption edge, but an iron
emission line is present at energy $\sim 6.6 $ keV (produced by Fe XXIII, 
Turner et al. 1992) and $\sim 6.7$ keV (Fe XXIV--Fe XXV),
respectively, implying a gradual increase of the ionization level in
the outflow. The equivalent width of the iron energy line at $\sim 6.6
$ keV is $\sim 30$ eV, that is smaller than the equivalent width of 
the iron emission line in the other spectra, probably because the
fluorescence yield associated to Fe XXIII is smaller than that 
associated to Fe XXIV and Fe XXV (Turner et al, 1992).  
In the hypothesis that the iron emission line is associated to 
the outflow, we deduce that also during intervals A2 and A3 the 
matter in the outflow is highly ionized. We argue that the outflow 
has a low density at these orbital phases because the absorption edge
is not observed (the corresponding equivalent hydrogen column is lower
than $10^{23}$ cm$^{-2}$).  The presence of the H${\alpha}$ line,
detected at these phases by Johnston et al. (1999), suggests that the
matter of the outflow is either photoionized or heated by a non
thermal emission. In fact, if the matter in the outflow would be
heated by a thermal emission, we should not observe an H${\alpha}$
emission, because recombination to neutral hydrogen should not be
possible.
 
In spectra A4 and B (orbital phase 0.05--0.19) an absorption edge is
present.  The energy of the edge is 8.46 keV and 8.71 keV,
respectively.  These energies are compatible with iron ionization
levels of Fe XXIII--XXV (Turner et {\it al.} 1992).  The
best fit value for the optical depth $\tau_{edge}$, considering the
photoionization cross section for the K--shell of Fe XXIII (Krolik \&
Kallman, 1987) and assuming cosmic abundance of iron, corresponds to
a hydrogen column density of $ \sim 6 \times 10^{23}$ cm$^{-2}$ and 
$\sim 7 \times 10^{23}$ cm$^{-2}$ for A4 and B respectively, in
agreement with the results reported by Iaria et al. (2001).  The iron
ionization level is similar to that in spectra A2 and A3
and the presence of the edge could be explained by an increase of the
density and/or optical depth of the outflow.  The iron emission line is
still observed in spectrum A4 (together with the absorption edge), but 
it is not detected is spectrum B.  

\subsubsection{The Quiet state}
  
The spectra C, D and E, corresponding to the orbital phase 0.3--0.6,
show a sub--Eddington flux: the lightcurve is quite constant at $\sim
300$ counts s$^{-1}$ and the hardness ratio increases from $\sim 0.75$
up to 0.9. 
The seed--photon radius is constant
at $\sim 90$ km, the seed--photon temperature is constant at $\sim
0.4$ keV, while the electron temperature increases from 1.2 keV up
to 1.5 keV.  In these spectra an absorption edge is present at $\sim
8.3$ keV, with optical depth decreasing from 0.12 to 0.06.  The
energy of $\sim 8.3$ keV indicates that the matter is less ionized and
the low values of the optical depth indicate that the absorption 
column decreases.  The
equivalent hydrogen column goes from $ \sim 2.6 \times 10^{23}$
cm$^{-2}$ down to $\sim 10^{23}$ cm$^{-2}$.  An iron emission line
is also present, the energy of which is stable at $\sim 6.7$ keV.  
This indicates that the outflow is still present, although it is now
probably much weaker.  

\subsubsection{The partial covering state: absorption from neutral matter 
coming from the companion star}
  
Cir X--1 has a highly eccentric orbit ($e \sim 0.9$; Murdin et al.,
1980; Tauris et al, 1999).  It is possible that, for the geometry of
the system (see e.g. Fig. 4 in Johnston et al., 1999), near the
periastron, the neutron star is occulted by the outer layers of the
companion-star atmosphere.  Also, the high eccentricity of the orbit
might produce large tidal interactions at these phases, which expand
and/or deform the stream of the accreting matter, maybe forming a bulge
in the outer accretion disk.  The excess of
neutral matter modifies the observed emission of the X-ray source,
producing a partial covering of the energy spectrum of
the source (see also Inoue 1989).  The partial covering component is
detected in spectra A1, F and G, corresponding to the phase interval
between 0.78 and 0.97.  In this state the (ionized) iron absorption edge 
and emission line are not detected.  At orbital phase $\sim 0.78$ the
hydrogen column, inferred by the partial covering, is $\sim 6.9 \times
10^{23}$ cm$^{-2}$, and it increases up to $\sim 11 \times 10^{23}$
cm$^{-2}$ at the orbital phase $\sim 0.89$.  This is probably due to
the approach of the X-ray source to the companion star.  At 0.95 
the hydrogen column decreases to $\sim 1 \times 10^{23}$
cm$^{-2}$, and the partial covering component disappears at following
orbital phases. The decrease of the hydrogen column can be caused 
by the increase of the accretion rate to super--Eddington values, 
which photoionizes the surrounding neutral matter.  
In fact the X-ray flux increases from
spectrum F to spectrum G (see Fig.~3), indicating that the accretion
rate increases at these phases.  

\subsection{Comparison with other luminous LMXBs}

Recently several Z--sources were analysed using a broad band 
(0.1--200 keV) energy
range (e.g. GX 17+2, Di Salvo et al., 2000; GX 349+2, Di Salvo et al.,
2001).  Their spectra show an absorption edge associated to highly
ionized matter, as it is observed in Cir X--1.  The hydrogen column
associated to the absorption edge in these Z--sources is $< 1 \times
10^{23}$ cm$^{-2}$ and the equivalent widths of the iron emission line
present in these spectra are $\sim 40-70$ eV. These equivalent widths,
according to Vrtilek et al. (1993) who suppose that emission lines are
produced in a photoionized ADC, could be expected in sources with
a low inclination angle.  
The phenomenon of the outflow from the accretion disk connected to
Eddington--sources is partially confirmed by the detection, in IR
band, of lines with P--Cygni profiles in Sco X--1 and GX 13+1 
(Bandyopadhyay et al., 1999). 
Because the outflow of ionized matter should be 
mostly along the disk surface, the absorption edge should be more
evident in sources with high inclination. Therefore the small values
of the optical depth of the absorption edge in GX 17+2 and GX 349+2
could be associated to low inclinations. In this case the
absorption edge might be considered an indicator of the
inclination angle of the source.

\section{Conclusions}

We have analyzed observations of Cir X--1 from ASCA archive, performed in
1998 March and corresponding to a whole orbital period.  
Cir X--1 is most probably a high inclination source, and this might explain 
why we do not observe the blackbody emission of the neutron star or inner
accretion disk. We observe a Comptonized component probably coming from a
ADC. We find that the optical depth of the Comptonizing region 
is $\sim 20$, the seed--photon temperature ranges from 0.4 to 0.5 keV, the 
electron temperature increases from 1 keV at phase 0.05 up to 1.8 keV at phase
0.9. The radius of the seed photons is around 90 km.  This could be
explained either by a magnetic field of $10^{10}$ G or, more probably,
by the presence of optically thick material in the inner part of the system 
which hides the innermost regions.  During the observations, which span an
entire orbital period of the X-ray source, it is possible to
distinguish three states of the source: a high-accretion rate (probably
super--Eddington) state, where
X--ray flares are present, near the periastron; a quiet state, where
the system accretes at a sub--Eddington rate, at orbital phases from
0.3 to $\sim 0.7$, and a partial covering state, at orbital
phases 0.78--0.97.
The first state is characterized by the presence of a prominent iron
absorption edge, probably produced in an outflow of highly ionized matter.  
The matter is swept away and photoionized by the high luminosity of the X-ray 
source in this state.
The second state shows an absorption edge of less ionized iron with an
equivalent hydrogen column smaller than $10^{23}$ cm$^{-2}$; the outflow is 
probably still present in this state, although much weaker. 
The third state is characterized by the presence of partial
covering, explained as due to absorption by the atmosphere of the companion 
star and/or by the accreting matter from the companion, which, at these orbital
phases, can fall in the line of sight  
between the X-ray source and the observer.

\acknowledgments 

This work was supported by the Italian Space Agency (ASI), by the
Ministero della Ricerca Scientifica e Tecnologica (MURST).  This
research has made use of data obtained through the High Energy
Astrophysics Science Archive Research Center Online Service, provided
by the NASA/Goddard Space Flight Center.

\clearpage

\section*{TABLES}
\begin{table}[th]
\begin{center}
\footnotesize

\caption{Log of the used ASCA observations.
For each of the analysed observations we show
the start and stop times (in Universal Time, second and third
column), the exposure time not corrected for the deadtime (fourth column),  
and the corresponding orbital phases (fifth column) referred to the
ephemeris obtained by Stewart et al. (1991).}
\label{tab1}
\vskip 0.5cm
\begin{tabular}{c|c|c|c|c}

\tableline  
\tableline
Observation  &Start--time & Stop--time& Exposure time (ks)  & Orbital Phase \\ 
\tableline 
A & 1998-03-03 12:12:49 UTC & 1998-03-05 18:50:33 UTC &38.2& 0--0.09, 
0.95--1\\
B & 1998-03-06 23:33:13 UTC   & 1998-03-07 09:00:43 UTC& 7.9 &0.17--0.19\\
C & 1998-03-09 04:19:38 UTC & 1998-03-09 15:10:39 UTC& 10.2&0.30--0.33\\
D & 1998-03-11 02:21:16 UTC & 1998-03-11 13:30:52 UTC& 6.9 &0.41--0.44\\
E & 1998-03-13 02:12:29 UTC & 1998-03-13 13:20:47 UTC  & 8.4 &0.54--0.56\\
F & 1998-03-17 00:44:30 UTC & 1998-03-17 11:20:24 UTC  & 4.9 &0.78--0.80\\
G & 1998-03-18 16:24:23 UTC   & 1998-03-19 00:21:19 UTC& 3.9 &0.87--0.89\\
\tableline
\end{tabular}
\end{center}
\end{table}

\begin{table}[th]
\begin{center}
\footnotesize
\caption{Spectral features, flux and luminosity in the selected
intervals.  For each of the intervals we report the spectral
features that were necessary to fit the corresponding spectra in addition
to the Comptonization model.
We also report the flux, in units of $10^{-8}$
ergs cm$^{-2}$ s$^{-1}$, and the luminosity, in units of $10^{38}$
ergs $s^{-1}$, assuming a distance of the source of 5.5 kpc (Case \&
Bhattacharya, 1998), calculated in the indicated energy ranges. }
\label{tab2}
\vskip 0.5cm
\begin{tabular}{c|c|c|c|c|c|c|c}

\tableline  
\tableline
Interval & Absorption & Partial& Emission & Flux& Flux &Luminosity& Orbital\\  
         & Edge       &Covering& Line     & 2.5--25 keV& 0.1--100 keV 
&0.1--100 keV& Phase  \\
\tableline 
A1 & No        &   Yes   &    No & 3.8  &    5.8 & 2.1  & 0.95\\
A2 & No        &   No    &   Yes & 3.8  &    5.4 & 1.9  & 0.99 \\
A3 & No        &   No    &   Yes & 4.1  &    7.0 & 2.5  & 0.01--0.04\\
A4 &Yes        &   No    &   Yes & 2.5  &    5.9 & 2.1  & 0.05--0.09 \\
B  &Yes        &   No    &    No & 2.8  &    6.4 & 2.3  & 0.17--0.19 \\
C  &Yes        &   No    &   Yes & 2.4  &    4.3 & 1.5  & 0.30--0.33\\
D  &Yes        &   No    &   Yes & 2.7  &    4.6 & 1.6  & 0.41--0.44\\
E  &Yes        &   No    &   Yes & 2.7  &    4.3 & 1.6  & 0.54--0.56\\
F  &No         &   Yes   &    No & 3.1  &    4.6 & 1.7  & 0.78--0.80\\
G  &No         &   Yes   &    No & 3.5  &    5.0 & 1.8  & 0.87--0.89\\
\tableline
\end{tabular}
\end{center}
\end{table}

\clearpage

\begin{table}[th]
\begin{center}
\footnotesize
\caption{Results of the fit of Cir X--1 for the first
 five spectra. The continuum model consists of a Comptonized
 spectrum modeled by Comptt. 
 Uncertainties are at 90\% confidence level for a
 single parameter.  $N_{\rm H_{PC}}$ is the hydrogen column
 corresponding to the partial covering, $f$ is the fraction of
 covered area by the neutral matter, $kT_0$ is the temperature of the
 seed photons for the Comptonization, $kT_e$ is the electron
 temperature, $\tau$ is the optical depth of the scattering cloud
 using a spherical geometry, R$_W$ is the radius of the seed-photon Wien
 spectrum in km, f$_{bol}$ is the intrinsic flux of the Comptonized
 component in units of ergs cm$^{-2}$ s$^{-1}$.  
 $EQW_{\rm Fe}$ is the equivalent width of the
 Gaussian emission line, E$_{Fe}$ the centroid energy of the 
 line and I$_{Fe}$ is the intensity of the emission line in
 units of photons cm$^{-2}$ s$^{-1}$.  F--Test indicates the
 probability of chance improvement of the fit when the feature 
 indicated in parentheses is included in the spectral model.}

\vskip 0.5cm
\begin{tabular}{l|c|c|c|c|c}

\tableline  
\tableline

Spectra    & A1 & A2 & A3 & A4 & B\\   
Orbital Phase  &  0.95  &  0.99  &  0.01--0.04   &  0.05--0.09  
             &  0.17--0.19 \\                           
\tableline                               
  &    &  &  & &   \\
$N_{\rm H}$ $\rm (\times 10^{22}\;cm^{-2})$&  1.630 (fixed) &  
$1.630^{+0.030}_{-0.033}$ & $1.6800^{+0.0094}_{-0.0078}$ & 
$1.5400^{+0.0084}_{-0.0067}$ & $1.627^{+0.011}_{-0.007}$\\

$N_{\rm H_{PC}}$ $\rm (\times 10^{23}\;cm^{-2})$ & $1.00^{+0.29}_{-0.46}$ &
          --  &--&--& --\\

$f$          &$0.21^{+0.12}_{-0.17}$& --&--&--& --\\

 E$_{edge}$ (keV)            & --& 8.3 (fixed) & 8.3 (fixed) & $8.464^{+0.102}_{-0.097}$ & 
$8.71^{+0.15}_{-0.16}$\\

$\tau_{\rm edge}$            & --& $<0.0094$   & $<0.037$    & $0.353 \pm 0.065$ &
$0.381 \pm 0.098 $\\

$k T_0$ (keV)                &  $0.507 \pm 0.033$ & 
$0.519^{+0.020}_{-0.007}$ & $0.4860^{+0.0025}_{-0.0020}$  &  
$0.4480 \pm 0.0017$ & $0.4403^{+0.0025}_{-0.0018}$ \\

$k T_{\rm e}$ (keV)     &  $1.556^{+0.136}_{-0.087}$ & $1.491^{+0.028}_{-0.011}$ &
 $1.2660^{+0.0030}_{-0.0019}$ & $1.0720^{+0.0026}_{-0.0016}$ & 
$1.1210 \pm 0.0027$\\

$\tau$                  &   $17.4 \pm 3.0$ & $20.07^{+0.23}_{-0.65}$ & 
$18.570^{+0.063}_{-0.048}$ & $16.120^{+0.055}_{-0.058}$ & 
$16.380^{+0.061}_{-0.099}$\\

 N$_{\rm comp}$                            &  $11.48^{+0.82}_{-1.17} $ & 
$10.91^{+0.18}_{-0.33}$ & $20.890^{+0.052}_{-0.119}$ & $27.020^{+0.094}_{-0.072}$ 
& $26.800^{+0.080}_{-0.123}$\\

f$_{bol}$($\times 10^{-8}$)     & 5.8 &5.4 & 7.0 & 5.9 & 6.4\\

R$_W$ (km)             & $71 \pm 10$ & $59.6 \pm 4.9$ & $87.8 \pm 1.2$ & 
                         $112.3 \pm 1.3$ &  $117.5 \pm 1.4$ \\

E$_{Fe}$ (keV)         & -- &  $6.616^{+0.038}_{-0.108}$ & 
$6.726 \pm 0.052$ & $6.731 \pm 0.068$ & --  \\

$\sigma_{\rm Fe}$ (keV)                    & -- & $< 0.17$ & 
$0.174^{+0.069}_{-0.072}$ & $< 0.15$& -- \\

I$_{\rm Fe}$ $(\times 10^{-2})$    & -- & $1.01^{+0.33}_{-0.24}$ & 
$1.84^{+0.31}_{-0.26}$ & $0.61^{+0.13}_{-0.12}$ & --\\

$EQW_{\rm Fe}$ (eV)            &-- & $33.2^{+10.8}_{-7.9}$ & 
$74.5^{+12.4}_{-10.6}$ & $67.0^{+14.7}_{-12.9}$ &-- \\

$\chi^2$/d.o.f.               & 309/391 & 252/389 & 244/389 & 259/387& 263/390\\
 
F--Test   (line)              & -- & $2.1 \times 10^{-28}$ & $\sim 0$  & $\sim 0$ 
                              & --  \\

F--Test   (edge)              & -- & $\sim 1$ & $\sim 1$ &$\sim 0$  & $4.17 \times 10^{-28}$ \\

F--Test   (PC)                & 0.08  &--& -- & --& --\\ 

\tableline
\end{tabular}
\end{center}
\end{table}

\begin{table}[th]
\begin{center}
\footnotesize
\caption{Results of the fit of Cir X--1 for the last five spectra.
See Table 3 for definitions.}

\vskip 0.5cm
\begin{tabular}{l|c|c|c|c|c}

\tableline  
\tableline

Spectra    & C & D & E & F & G\\ 
 Orbital Phase  &    0.30--0.33&   0.41--0.44& 0.54--0.56&   0.78--0.80& 
            0.87--0.89\\               
\tableline                               
  &    &  &  & &   \\
$N_{\rm H}$ $\rm (\times 10^{22}\;cm^{-2})$&  $1.542^{+0.046}_{-0.017}$ &  
$1.629^{+0.035}_{-0.068}$ & $1.628 \pm 0.054$ & 
$1.619 \pm 0.080$ & $1.659^{+0.080}_{-0.097}$\\

$N_{\rm H_{PC}}$ $\rm (\times 10^{23}\;cm^{-2})$ &--&--&--& $6.9^{+3.6}_{-2.4}$&
 $10.8^{+8.4}_{-6.8}$\\

$f$          & --&--&--& $0.207^{+0.106}_{-0.067}$& $0.25^{+0.66}_{-0.15}$\\

 E$_{edge}$ (keV)         & $8.23 \pm 0.11$ & 8.3 (fixed) & 
                         $8.34^{+0.31}_{-0.27}$ & --&--\\

$\tau_{\rm edge}$         &$0.151^{+0.030}_{-0.039}$ & $< 0.075$ &
$0.062^{+0.046}_{-0.033}$ & --& --\\

$k T_0$ (keV)                &  $0.396^{+0.013}_{-0.016}$ & 
$0.416^{+0.026}_{-0.011}$ & $0.391 \pm 0.020$  &  
$0.383^{+0.029}_{-0.034}$ & $0.369^{+0.041}_{-0.017}$ \\

$k T_{\rm e}$ (keV)     &  $1.280^{+0.022}_{-0.018}$ & $1.354^{+0.027}_{-0.011}$ &
 $1.505^{+0.032}_{-0.018}$ & $1.699^{+0.013}_{-0.035}$ & 
$1.762^{+0.031}_{-0.034}$\\

$\tau$               &    $20.92^{+0.33}_{-0.48}$&   $19.48^{+0.25}_{-0.60}$ & 
$19.94^{+0.29}_{-0.48}$ & $20.17^{+0.54}_{-0.41}$ & $20.16^{+0.45}_{-0.32}$\\

 N$_{\rm comp}$             &   $12.29 \pm0.37$&  $12.28^{+0.26}_{-0.72} $ 
& $9.46^{+0.31}_{-0.47}$ & $7.80 \pm 0.78$  & $8.1^{+4.6}_{-1.7}$\\

f$_{bol}$($\times 10^{-8}$)     & 4.3 & 4.6 & 4.3 & 4.6 & 5.0\\ 

R$_W$ (km)             & $93 \pm 8$ & $ 91 \pm 12$ & $ 94 \pm 10$ & 
                         $95 \pm 17$ &  $106 \pm 23$ \\

E$_{Fe}$ (keV)         &$6.759^{+0.074}_{-0.064}$  &  $6.811^{+0.095}_{-0.100}$ & 
$6.763 \pm 0.099$ & -- & --  \\

$\sigma_{\rm Fe}$ (keV)            & $0.196^{+0.122}_{-0.077}$  & $0.25 \pm 0.16$& 
$0.27^{+0.16}_{-0.19}$ & --& -- \\

I$_{\rm Fe}$ $(\times 10^{-2})$  & $1.05^{+0.33}_{-0.20}$  & 
$1.16^{+0.43}_{-0.37}$ & 
$1.12^{+0.43}_{-0.46}$ & --& --\\

$EQW_{\rm Fe}$ (eV)            & $63.0^{+19.5}_{-12.2}$ & 
   $70.7^{+26.5}_{-22.4}$ & $50.9^{+19.7}_{-20.8}$ & --&-- \\

$\chi^2$/d.o.f.               & 256/387 & 251/387 & 250/387 & 285/390& 309/390\\
 
F--Test   (line)              &  $\sim 0$  & $3.6 \times 10^{-38}$ & 
                            $8.0 \times 10^{-10}$  & -- & --  \\

F--Test   (edge)              & $1.2 \times 10^{-7}$  & $\sim 1$ &
                              $6.2 \times 10^{-22}$  & -- & -- \\

F--Test   (PC)              & --& -- & --& $2.8 \times 10^{-10}$  & 
 $5.3 \times 10^{-5}$ \\

\tableline
\end{tabular}
\end{center}
\end{table}

\clearpage
 
\section*{FIGURE CAPTIONS}
\bigskip

\noindent
{\bf Figure 1}: Upper panel: Cir X--1 lightcurve in the energy band
0.6--10 keV. The intervals in which we divided these observations are also 
indicated with the corresponding labels. 
Lower panel: The ratio of the count rate in the energy
band 3--7 keV to that in band 1--3 keV.  The bin time is 5595 s.\\ 
{\bf Figure 2}: The best fit values of the seed--photon
temperature, electron temperature and radius of the seed photons
are plotted versus the orbital phase of Cir X--1 in the upper,
middle and lower panel, respectively. \\
{\bf Figure 3}: Unabsorbed flux of Cir X--1 extrapolated in the band 
0.1--100~keV. We marked with a solid line the flux of $\sim 5 \times 
10^{-8}$ erg s$^{-1}$ cm$^{-2}$, which is the Eddington flux for a
neutron star with a mass of 1.4$M_{\odot}$ at a distance of 5.5 kpc.

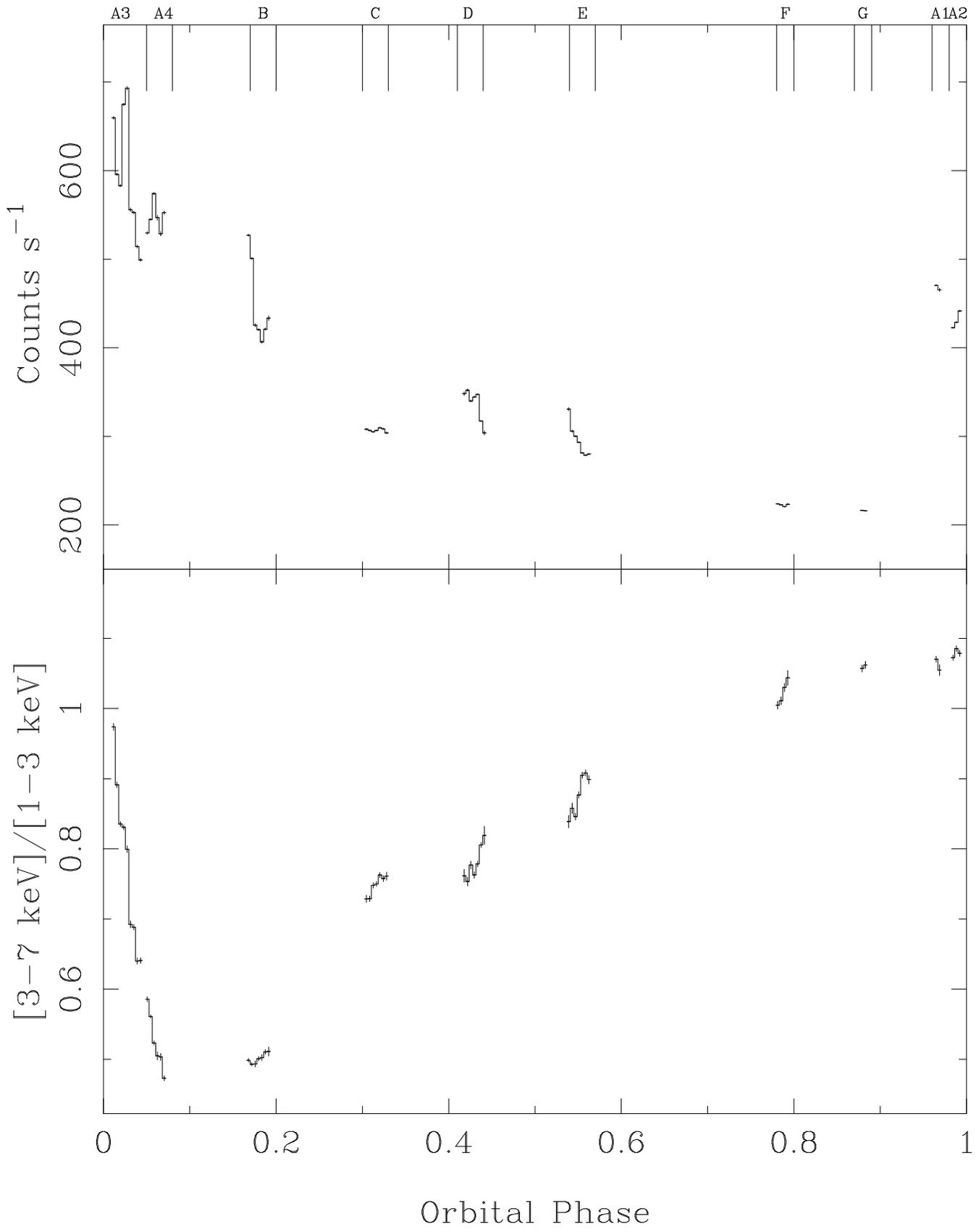
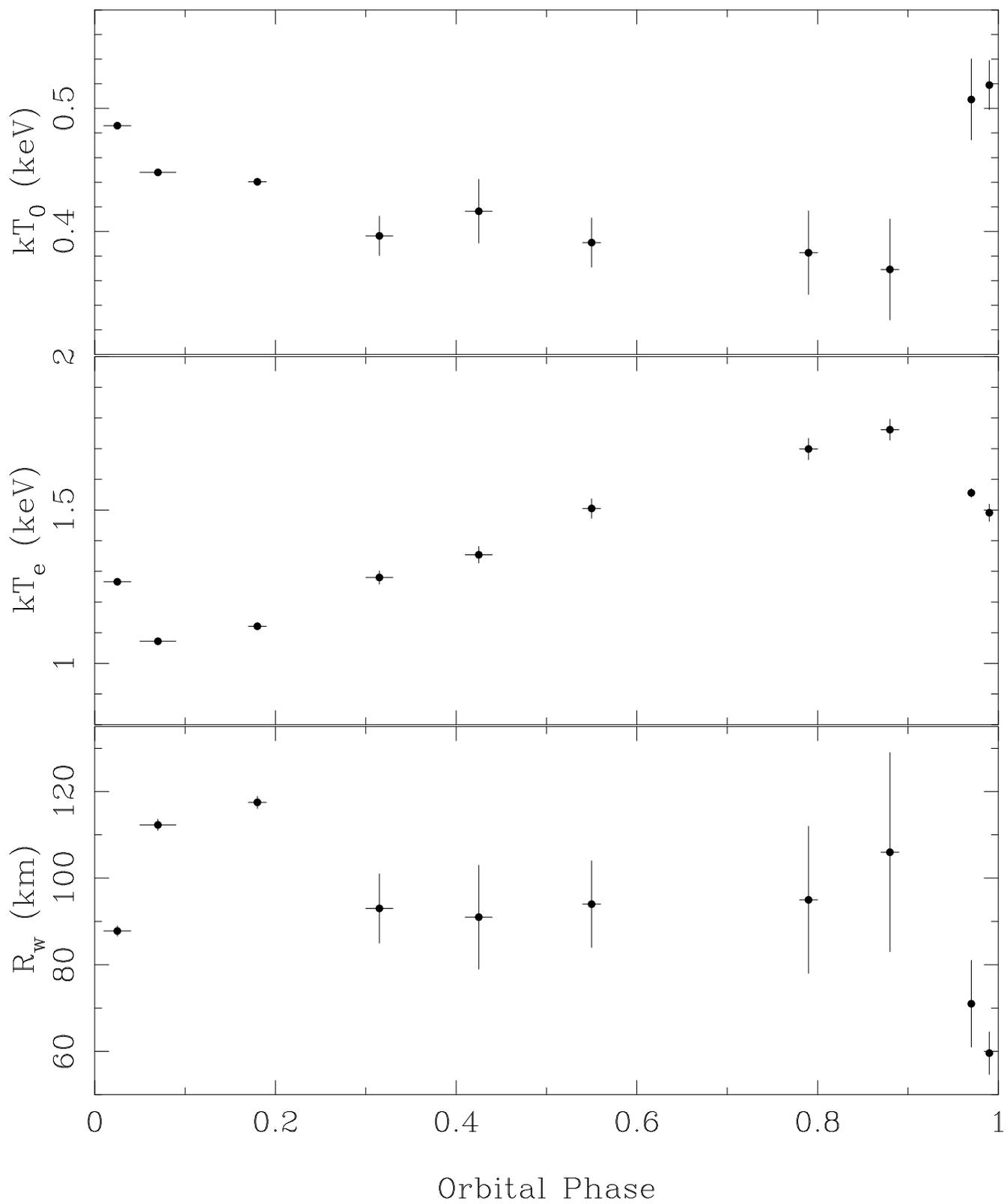
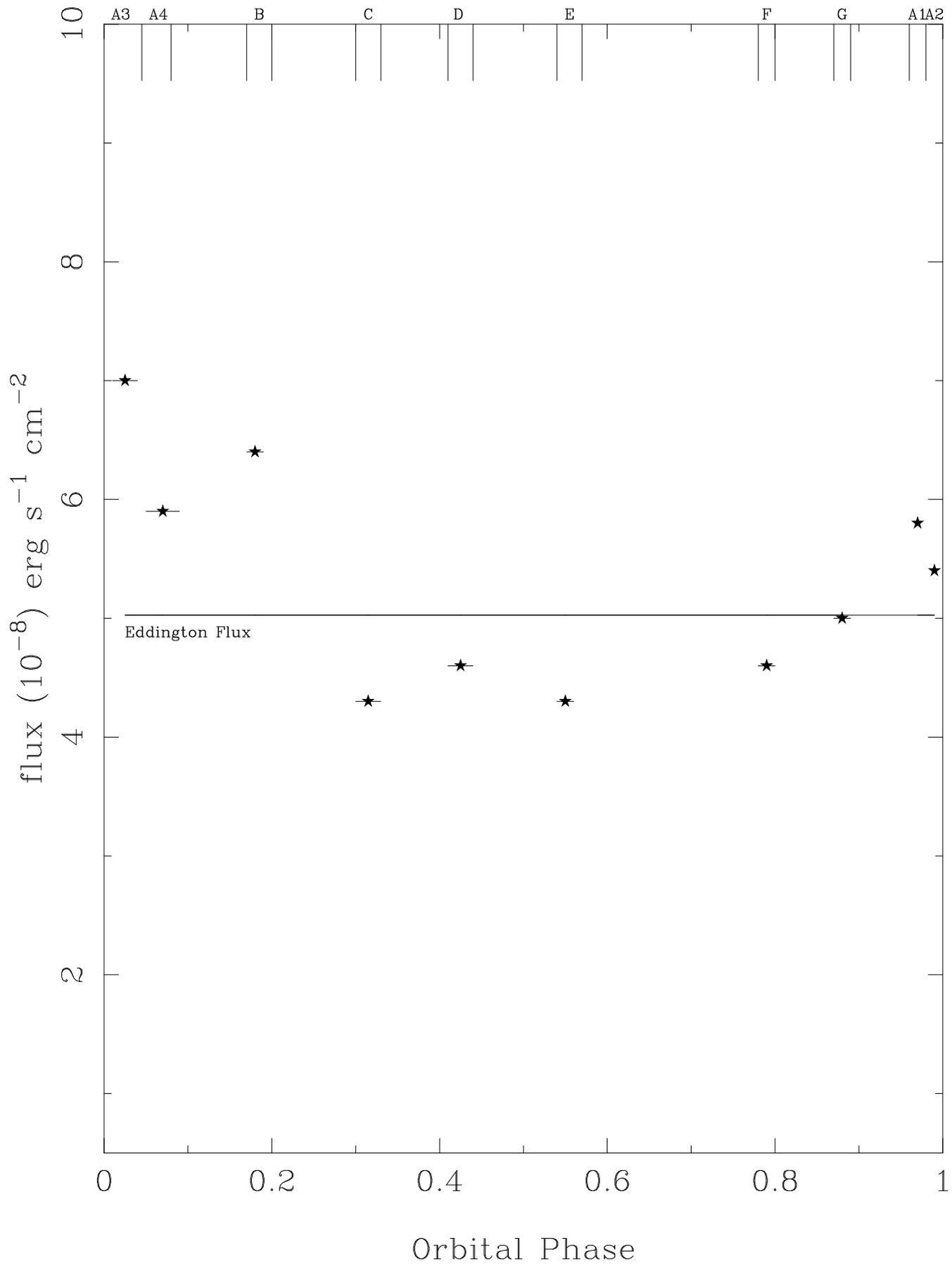
\end{document}